   \documentstyle[12pt,epsf]{article}

\def\thefootnote{\alph{footnote}}
\def\thefootnote{\fnsymbol{footnote}}
\newcommand{\eq}{\begin{equation}}
\newcommand{\en}{\end{equation}}
\newcommand{\eqa}{\begin{eqnarray}}
\newcommand{\ena}{\end{eqnarray}}

\newcommand{\lbl}{\label}

\def\real{{\rm Re}\hskip 1pt}
\def\ee#1{{\rm e}^{#1}}
\def\trace{{\rm Tr}\hskip 1pt}
\def\ii{{\rm i}}
\def\diag{{\rm diag}}

\def\cJ{{\cal J}}
\def\mycaptionl#1#2{%
\begin{center}
\hskip 1pt\vskip -1cm
\begin{minipage}{14cm}
\small {\bf #1}: {\sl #2}
\end{minipage}
\null\hskip 1pt\vskip -0.2cm
\end{center}}
\newcommand{\NP}[1]{Nucl.\ Phys.\ {\bf #1}}
\newcommand{\PL}[1]{Phys.\ Lett.\ {\bf #1}}

\begin{document}
\hskip 10cm \vbox{
\hbox{DFTT 6/97}
\hbox{NORDITA 97/5P}
\hbox{\tt hep-th/9701145}}
\vskip 0.8cm
\centerline{\Large\bf Analytic results in 2+1-dimensional}
\centerline{\Large\bf Finite Temperature LGT}
\vskip 0.8cm
\centerline{\large M. Bill\'{o}\footnote{E--mail: {\tt billo@alf.nbi.dk};
supported by I.N.F.N., Italy}}
\vskip .2cm
\centerline{\small\it Nordita, Blegdamsvej 17, Copenhagen \O, Denmark}
\vskip .4cm
\centerline{\large M. Caselle\footnote{E--mail: {\tt caselle@to.infn.it}},
A. D'Adda}
\vskip .2cm
\centerline{\small\it Istituto Nazionale di Fisica Nucleare, Sezione di Torino}
\centerline{\small\it  Dipartimento di Fisica
Teorica dell'Universit\`a di Torino}
\centerline{\small\it  via P.Giuria 1, I-10125 Turin,Italy}
\vskip 1.2cm
\begin{center}
{\bf Abstract}\\
\vskip 0.5cm
\begin{minipage}{14cm}
\small 
In a 2+1-dimensional pure LGT at finite temperature the critical coupling
for the deconfinement transition  scales as $\beta_c(n_t) = J_c n_t + a_1$, 
where $n_t$ is the number of links in the ``time-like'' direction of the 
symmetric lattice.
We study the effective action for the Polyakov loop obtained by neglecting 
the space-like plaquettes, and we are able to compute analytically in this
context the coefficient $a_1$ for any SU$(N)$ gauge group;
the value of $J_c$ is instead obtained from the effective action by means
of (improved) mean field techniques. Both  coefficients have already been
calculated in the large $N$ limit in a previous paper. 
The results are in very good agreement with the existing Monte Carlo 
simulations. This fact supports the conjecture that, in the 2+1-dimensional 
theory, space-like plaquettes have little influence on the dynamics of the
Polyakov loops in the deconfined phase.  
\end{minipage}
\end{center}
\noindent
\setcounter{footnote}{0}
\def\thefootnote{\arabic{footnote}}
\section{Introduction and setting}
\label{intro}
Let us consider a finite temperature lattice gauge theory  (LGT) with 
gauge  group SU$(N)$, defined on a $2+1$ dimensional cubic lattice.
The ``time'' dimension is compactified with
periodic boundary conditions and compactification length 
${1 \over T} = N_t a_t$, where $N_t$ is the number of lattice spacings 
in the time direction, and $a_t$ the corresponding lattice spacing;
$T$ is the temperature. 
We consider in general asymmetric lattices; 
denoting by $a_s$ the 
lattice spacing in the space directions, the asymmetry parameter $\rho$
is defined as:
\eq
\rho = { a_{s} \over a_t}= T N_t a_s~.
\label{rho}
\en
Moreover we introduce different bare couplings $\beta_t$ and $\beta_s$ in 
the time and space directions. 
The Wilson action is then\footnote{The notations throughout this paper
are the same as the ones of Ref. ~\cite{review}, to which we refer for a 
more detailed exposition of the general setting of the theory. 
In particular we choose the normalization of the couplings 
$\beta_s$ and $\beta_t$ so as to have a smooth large $N$
limit, since we will compare the results at finite $N$  
with the ones obtained in the large $N$ limit.}
\eq
S_{\rm W}= N \sum_{\vec{x},t}~ \left(\sum_{i=1}^2\beta_t 
\real\trace G_{0i}(\vec{x},t)+ \beta_s  
\real\trace G_{12}(\vec{x},t)\right)~,
\label{wilson}
\en
where $G_{0i}$ and $G_{12}$ are the time-like and space-like plaquette
variables:
\begin{eqnarray}
G_{0i}(\vec{x},t)& = &   V(\vec{x},t)U_i
(\vec{x},t+1) V^{\dagger}(\vec{x}+\hat{\i},t)U^{\dagger}_i(\vec{x},t)~,
 \nonumber \\ G_{12}(\vec{x},t)& = &   U_1(\vec{x},t)U_2(\vec{x}+
\hat{\i},t)U^{\dagger}_1(\vec{x}+ \hat{\j},t)U^{\dagger}_2(\vec{x},t)~.
\label{plaq}
\end{eqnarray}
We have denoted by $U_i(\vec{x},t)$ the link 
variables in the space directions and by 
$V(\vec{x},t)$ the link variables in the time direction. 

For a given $\rho$ the relation between the couplings $\beta_t$ and 
$\beta_s$ can be obtained by requiring that the Wilson action 
(\ref{wilson}) reproduces in the continuum limit the usual Yang--Mills
action.
At the classical level, this leads to the relations
\eq
\frac{2}{Ng^2}=a_s\sqrt{\beta_s\beta_t}~,
\hskip 1cm
\rho = \sqrt{\frac{\beta_t}{\beta_s}}~,
\label{couplings}
\en
which show that
in $d=2$ the coupling constant $g^2$ has the dimensions of a mass.
This is the main difference with the $d=3$ case and it is the reason
for which the $d=2$ models are much easier to study.
In fact in this case the coupling constant itself sets the 
overall mass scale. Near the continuum limit a physical observable, 
like the critical temperature $T_c$, with the dimensions of a mass can be
written, according to the renormalization group equations, as a series in 
powers of $g^2$. Hence in terms of the coupling\footnote{It is not
mandatory to measure $T_c$ in units of $a_t$ and to choose $\beta_t$ instead of
$\beta_s$ in eq. (\ref{251}), but these choices simplify the 
comparison with the results of Sec. \ref{conto}.}
$\beta_t$ we have
\eq
a_tT_c=\frac{c_1}{\beta_t}+\frac{c_2}{\beta_t^2}+\cdots~,
\label{251}
\en
with $c_1$ and $c_2$ suitable constants.
The critical temperature is obtained by looking at the critical coupling
$\beta_c$ at which the deconfinement transition occurs. Since
 the lattice size in the time direction is $N_t\equiv\frac{1}{Ta_t}$, 
we can rephrase eq. (\ref{251}) as a scaling law for the 
critical coupling $\beta_c$ as a function of $N_t$: 
\eq
\beta_c(N_t)=J_c N_t+  a_1 \ldots~,
\label{252}
\en
where we have introduced the two constants $J_c\equiv c_1$ and 
$a_1\equiv \frac{c_2}{c_1}$ for future convenience. 
Finding the precise values of these two
constants is one of the aims of the present paper.

It is clear from eqs. (\ref{rho}) and (\ref{couplings}) that equivalent 
regularizations with different values of $\rho$ require different
values of $N_t$. Hence, to maintain the  equivalence, $N_t$ must be
a function of $\rho$ according to eq. (\ref{rho}). 

Among all these equivalent regularizations a particular role is played 
by the symmetric one, which is defined by
\eq
\beta\equiv\frac{2}{aNg^2} 
\label{coupsym}
\en
(from now on we shall distinguish the symmetric regularization from the
asymmetric ones by eliminating the subscripts $t$ and $s$ in $\beta$ and
$a$).
By comparing eq.s (\ref{rho},\ref{couplings},\ref{coupsym}) we see that 
all regularizations are equivalent to the symmetric one provided
\eq
\beta=\rho\beta_s=\frac{\beta_t}{\rho}~,
\label{rel1}
\en
\eq
N_t(\rho)=\rho ~n_t~,
\label{rel2}
\en
where $n_t$ is the number of links in the time direction with a symmetric
regularization: $n_t = N_t(\rho=1)$.

Notice however that these equivalence relations have been derived in the 
{\em naive} or ``classical'' continuum limit, and quantum corrections 
are in general present. 
They lead in the large-$\rho$ limit to the following expressions:
\eq
\beta_t=\rho(\beta+\alpha^0_\tau)+ \alpha^1_\tau+\ldots~,
\hskip 1cm
\beta_s=\frac{\beta+\alpha^0_\sigma}{\rho}+\frac{\alpha^1_\sigma}{\rho^2}
+\ldots~.
\label{rel4}
\en
The constants $\alpha^{0,1}_{\tau\sigma}$
encode the quantum corrections to eq. (\ref{rel1}).
These corrections were studied and calculated in
the (3+1) dimensional case  by F. Karsch in~\cite{k81}.  

It is well known that, as a consequence of the periodic boundary conditions 
in the time direction, the Wilson action enjoys a new ${\bf Z}_N$ global
symmetry. 
Moreover, it becomes possible to define gauge invariant observables which are 
topologically non-trivial, such as the Polyakov 
loop\footnote{We will refer to the untraced quantity $P(\vec{x})$ as the 
Polyakov line.} 
\eq
\hat P(\vec{ x})=\trace P(\vec{x})= {\rm Tr}
 \prod_{t=1}^{N_t} V(\vec x,t)~.
\label{polya}
\en
The Polyakov loop is a natural order parameter for the ${\bf Z}_N$ symmetry.
In the
high temperature, deconfined phase ($T>T_c$) we have 
$\langle\hat{P}(\vec x)\rangle\not= 0$, while in the low
temperature, confining domain ($T<T_c$), the symmetry is unbroken, namely 
$\langle\hat{P}(\vec x)\rangle = 0$.

The way to approach the  description of the deconfinement transition
is to construct an effective action for the relevant degrees of
freedom, 
namely the Polyakov loops. This would require to perform the exact integration
over all the gauge degrees of freedom on the space-like links, a clearly
impossible task. In \cite{review,su2} we discussed a perturbative approach,
in which the effective action for the Polyakov loop is calculated exactly
order by order in an expansion in powers of $\beta_s$. Except for the theory
with $N=2$ ( see \cite{su2}) only the zeroth order of this expansion, which
corresponds to neglecting the space-like plaquettes altogether, has been
studied and used to determine the critical temperature. 
The point is that the approximation of throwing away the space-like plaquettes
appears to be a rather crude one if one wants to go beyond purely qualitative
results.  It is true however that far away from the continuum limit, that is
for very small $N_t$, the deconfinement transition occurs for small values 
of $\beta$ (we refer here to a symmetric lattice) and that the couplings among
Polyakov loops induced by the space-like plaquettes, beeing of higher order 
in $\beta$, can be safely neglected.  As we proceed towards the continuum limit
(large $\beta$) there is no obvious reason why the space-like plaquettes should
be uninfluential. In fact we know that in $3+1$ dimensions the effect of the
space-like plaquettes is crucial, as the rescaling of the critical 
coupling with $N_t$, which is linear in their absence, 
becomes logarithmic, according to 
asymptotic freedom, if they are taken into account.
The situation is quite different in $2+1$ dimensions, where the
rescaling of the critical coupling is linear both with and without 
space-like plaquettes. This is clearly related to the fact that the 
dimensional coupling constant sets the mass scale of the theory.
So in $2+1$ dimensions we are in a situation where the zeroth order
approximation is on one hand a good approximation of the full theory
for very small $N_t$, and on the other hand it has the same scaling properties
of the full theory near the continuum limit.
It is natural to conjecture at this point that the effective action for the
Polyakov loop obtained by neglecting the space-like plaquettes is a good
approximation even in the continuum limit and up to the temperature where the
deconfinement transition occurs.
Checking this conjecture by comparing the analytic predictions based on the
zeroth order approximation with the Monte Carlo simulations based on the full
theory is the purpose of this paper.

This paper is organized as follows. In the next section we shall construct
the effective action for the Polyakov loop  mentioned above. This section 
contains material which has been already presented elsewhere~\cite{review}, and
has been inserted so as to make this paper self-contained. In Sec. 3 we shall
obtain in the framework of the $\beta_s=0$ approximation the scaling
law (\ref{252}) and the explicit value of $a_1$ coefficient. 
The exact calculation of $a_1$ is the main new analytic result of the paper. 
In the last section we shall estimate by means of suitably improved mean field 
techniques the $J_c$ coefficient and compare our results with those
obtained with Monte Carlo simulations. 
The agreement of the analytic predictions with the 
Monte Carlo simulations, which seems to confirm the above conjecture, is 
the main physical result of our analysis. 
\section{The effective action}
\label{sez1} 
In order to derive the effective action for the Polyakov loops at $\beta_s=0$,
one starts from the time-like part of the Wilson action (\ref{wilson})
and performs the character expansion\footnote{For the derivation of this and of
the following formulas, see for instance Ref. ~\cite{drouzub}.} 
\eq
\label{fullac}
\ee{S_{\rm W}(\beta_s=0)} = \prod_{\vec{x},t,i} \left\{\sum_r d_r D_r(\beta_t) 
\chi_r (G_{0i}(\vec{x},t))\right\}~.
\en
In the above equation $d_r$ is the dimension of the representation
$r$ and  $\chi_r(U)$ the character of a matrix $U$ in this representation;
the coefficients $D_r(\beta_t)$ are given by
\eq
\label{dierre}
D_r(\beta)=\frac{F_r(\beta)}{d_r F_0(\beta)}~,
\en
with
\eq
\label{ferre}
F_r(\beta)\equiv
\int DU\, {\rm e}^{N\beta {\rm Re} {\rm Tr} U}\chi^*_r(U)
= \sum_{n=-\infty}^{\infty} {\rm det}\,I_{r_j+i+n}(N\beta)~.
\en
The $r_j$'s are a set of integers labelling the representation 
$r$ and they are constrained by $r_1>\cdots> r_N=0$.
The indices $1\leq i,j \leq N$ label the entries of the $N\times N$ matrix 
the determinant of which is taken and $I_n(\beta)$ denotes the modified 
Bessel function of order $n$.

As a result of neglecting the space-like plaquettes, each space link
variable appears only in two time-like plaquettes, and the integration
over the variables $U_i(\vec x,t)$ can be easily performed by  using the rules 
for the integration of characters (see \cite{review} for the details). 
One gets the following effective action for the Polyakov loops:
\eq
e^{S_{\rm Pol}(\beta_s=0)} = \prod_{\vec{x},i}
\sum_r  \left[ D_r(\beta_t) \right]^{N_t} \chi_r(P(\vec{x}+\hat\i)) 
\chi_r(P^\dagger(\vec{x})) ~.  
\label{zeta1}
\en
{}From the effective action we obtain some general informations on
the scaling behaviour that we may expect for the critical coupling.
They are obtained by requiring that the physics is unaffected by
changes of $N_t$ provided $\beta_t$ is suitably rescaled. In particular
it is easy to derive that the rescaling of $\beta_t$ is linear.

In fact if we just assume that $\beta_t$ becomes large for large $N_t$,
we can replace in (\ref{zeta1}) the coefficients $D_r(\beta_t)$ 
with their large $\beta_t$ behaviour which is well known and is given by 
\eq
\lim_{\beta_t\to\infty} D_r(\beta_t)={\rm e}^{-\frac{C_r}{2N\beta_t}+\ldots}~,
\label{242}
\en
where $C_r$ denotes the quadratic Casimir in the $r^{\rm th}$ 
representation:
\eq
\label{casimiro}
C_r = \sum_{i=1}^N r_i^2 - \sum_{i=1}^N r_i - {N(N^2-1)\over 12}~.
\en
So we have at the leading order
\eq
\label{largebeta}
\left[ D_r(\beta_t) \right]^{N_t} 
\stackrel{\beta_t \rightarrow\infty}{\longrightarrow} 
\exp \left(- {C_r N_t  \over 2 N \beta_t(N_t)}\right). 
\en
Clearly the above discussion is independent from the asymmetry of the
lattice, and in particular the case of the symmetric lattice can be recovered
by simply replacing $\beta_t$ and $N_t$ with $\beta$ and $n_t$ respectively.
The r.h.s. of (\ref{largebeta}), and hence the effective action (\ref{zeta1}),
are independent from $N_t$ in the large $N_t$ limit provided $\beta_t(N_t)$
depends linearly from $N_t$.
If we define, consistently with eq. (\ref{252}),
\eq
\label{jdef}
J = \lim_{N_t \to \infty} \frac{\beta_t(N_t)}{N_t}
\en
we obtain the large $N_t$ limit\footnote{The large $N_t$ limit can correspond 
either to the continuum limit, for instance in the symmetric lattice where 
$\rho=1$ and $N_t=n_t$, or to the limit $\rho \to \infty$, which is a continuum
limit only in the time dimension. These two limits, although conceptually
completely different, produce here the same effect as a consequence of having
neglected the space-like plaquettes.} of the effective action (\ref{zeta1}):
\eq
\label{zetah1}
e^{S_{\rm Pol}^{\rm hk}}  = \prod_{\vec{x},\hat\i}
\sum_r  \chi_r(P(\vec{x}+\hat\i)) \chi_r(P^\dagger(\vec{x})) 
\exp \left(- {C_r \over 2 N J}\right)~.
\en
The above effective action is the same one that we would have obtained,
for any value of $N_t$ and not just in the large $N_t$ limit,
if we had started, always within the $\beta_s = 0$ approximation, from the 
heat kernel action instead of the Wilson action
and with a coupling $\beta_{\rm hk}(N_t)$ given by
\eq
\label{betahk}
\beta_{{\rm hk}}(N_t) = J N_t~.
\en 

The critical value of $J$ at the deconfinement transition will be
determined in Sec. \ref{mf} by using an improved mean field method
and a Poisson resummation of eq. (\ref{zetah1}).
In the next section instead we shall calculate the next to leading term
in the rescaling of $\beta_t$, namely the coefficient $a_1$ of eq. (\ref{252}).
Unlike $J$, this depends on the particular form of the action on the lattice
(it is zero if we start from the heat kernel action as shown in (\ref{betahk}))
and it is completely determined by the form of the coefficients $D_r$.
\section{The scaling law: exact calculation of the subleading term}
\label{conto}
We have shown in the last section that the effective action  (\ref{zeta1})
has a smooth large $N_t$ limit, given in eq. (\ref{zetah1}), provided a 
linear rescaling of $\beta_t$ is assumed. 
In this section we want to go further, and assuming a linear riscaling 
of the form
\eq
\label{rescaling}
\beta_t(N_t) =  J N_t + a_1
\en
we shall determine the next to leading coefficient $a_1$ by requiring 
that the effective action is independent from $N_t$ also in the next to 
leading order of its $1 \over N_t$ expansion.
More precisely we shall show that it exists, for each $N$, a value of $a_1$,
{\it independent from the representation } $r$,
for which $(D_r(\beta_t))^{N_t}$ is independent from $N_t$ in the large
$N_t$ limit up to order ${1 \over N_t^2}$.

In order to do that we need to find the large $\beta_t$ asymptotic
expansion of $D_r(\beta_t)$ up to the order ${1 \over \beta_t^2}$,
namely one order higher than the leading one, already given in 
eq. (\ref{242}) and used in the derivation of the heat kernel action 
(\ref{zetah1}).

To begin with let us write $F_r(\beta)$, defined in (\ref{ferre}), as
\eq
\label{frint}
F_r(\beta) = \sum_{n=-\infty}^{\infty} \sum_P (-1)^{\sigma(P)} 
\int_{-\pi}^{\pi} 
\prod_i d \theta_i\, \ee{\ii \sum_j (r_j+P(j)+n)\theta_j + N\beta  
\sum_j \cos \theta_j}~.
\en
The infinite sum over $n$ implies the costraint $\sum_i \theta_i=0$; 
as a consequence a shift  of {\it all} the integers $r_j$ by an arbitrary, not
necessarily integer, quantity does not affect the r.h.s. of (\ref{frint}).
We can therefore replace the $r_j$'s in (\ref{frint}) with 
\eq
\hat{r}_j = r_j - \frac{1}{N} \sum_{k=1}^{N} r_k~,
\label{errehat}
\en
which satisfy the relation
\eq
\sum_j \hat{r}_j =0~.
\label{sumzero}
\en
We then proceed to the following redefinitions:
\eq
\label{chvar}
\xi_j = \sqrt{N \beta}~ \theta_j,~~~~~~~~~\nu_j = 
\frac{\hat{r}_j}{\sqrt{N \beta}},
~~~~~~~~~~~\nu = \frac{n}{\sqrt{N \beta}}.
\en
With these replacements eq. (\ref{frint}) can be rewritten as
\eq
\label{frint2}
F_r(\beta) = \sum_{\nu} \int_{-\pi \sqrt{N \beta}}^{\pi \sqrt{N \beta}}
\prod_i d \xi_i\, \cJ(\frac{\xi}{\sqrt{N \beta}})\, \ee{\ii 
\sum_j (\nu_j+\nu) \xi_j
+ N \beta \sum_j \cos\frac{\xi_j}{\sqrt{N \beta}} }~,
\en
where $\cJ(\frac{\xi}{\sqrt{N \beta}})$ is the unitary 
Vandermonde determinant defined by
\eq
{\cal J}(\theta) = \prod_{i<j} 2 \sin \frac{\theta_i - \theta_j}{2}~.
\lbl{vandermonde}
\en

We shall now derive the asymptotic expansion in $\frac{1}{\beta}$ of
(\ref{frint2}), {\it keeping }$\nu_j$ {\it fixed}, in spite of the
$\beta $ dependence hidden in the definition of $\nu_j$. Only at the end of
the calculation we shall replace $\nu_j$ with its expression.

The asymptotic expansion of the integral at the r.h.s. of (\ref{frint2})
is obtained in two steps. First, by noticing that under the integral
$\xi_j = -\ii \frac{\partial}{\partial \nu_j}$, we extract the Vandermonde
determinant and write
\eq
\label{frint3}
F_r(\beta) = \cJ(\frac{-\ii}{\sqrt{N \beta}} \frac{\partial}{\partial \nu})
\sum_{\nu} \int_{-\pi \sqrt{N \beta}}^{\pi \sqrt{N \beta}} \prod_i d \xi_i 
\,\ee{\ii \sum_j (\nu_j+\nu) \xi_j + N \beta \sum_j 
\cos\frac{\xi_j}{\sqrt{N \beta}} }~.
\en
The asymptotic behaviour of the above integral can now be easily calculated by 
expanding the cosine and using the saddle point method. Also, the integrals
over the $\xi_i$'s can be taken from $-\infty$ to $+\infty$ up to terms
which are exponentially depressed. The result is:
\eqa
\label{frint4}
F_r(\beta) & = & \cJ(\frac{-\ii}{\sqrt{N \beta}} \frac{\partial}{\partial \nu})
\sum_{\nu}  \ee{-{1 \over 2} \sum_j (\nu_j + \nu)^2} \nonumber\cr
& \times & \left[ 1 + \frac{1}{8 N 
\beta} (1 - 2 \sum_j (\nu_j + \nu)^2 + \frac{1}{3} \sum_j (\nu_j + \nu)^4 )+
C + O(\frac{1}{N^2 \beta^2}) \right]~,
\ena
where $C$ collects all terms that do not depend on the representation, that is
that are independent from $\nu_j$. 
These are in fact irrelevant for our calculation
as only the ratio  $\frac{F_r(\beta)}{F_0(\beta)}$ appears in (\ref{dierre}).
Similarly the terms of order $\frac{1}{N^2 \beta^2}$ have not been evaluated
because they are either representation independent or of order higher than
$\frac{1}{\beta^2}$ once the $\beta$ dependence of $\nu_j$ is 
taken into account.
  
The next step is the summation over $\nu$. It is easy to see that up to non
perturbative terms we can perform the replacement
\eq
\label{replacement}
\sum_{\nu} \rightarrow \sqrt{N \beta} \int_{-\infty}^{\infty} d \nu
\en
and that the resulting gaussian integrals can be calculated. With the condition
$\sum_j \nu_j=0$, the result is:
\eq
\label{frint5}
F_r(\beta) = \cJ(\frac{-\ii}{\sqrt{N \beta}} \frac{\partial}{\partial \nu})
\,\ee{-{1 \over 2} \sum_j \nu_j^2} \left[ 1 - \frac{N-1}{4\beta N}
\sum_j \nu_j^2 + \frac{1}{24 \beta} \sum_j \nu_j^4 +C +O(\frac{1}{N^2 
\beta^2}) \right]~.
\en
We have now to perform the derivatives included in the unitary Vandermonde determinant $\cJ$.
By expanding the sines (see eq. (\ref{vandermonde})) we can first write
it, up to the relevant order in $\frac{1}{\beta}$,
as
\eq
\label{jvsdelta}
\cJ(\frac{-\ii}{\sqrt{N \beta}} \frac{\partial}{\partial \nu}) =
\left(1 + \frac{1}{48 \beta} \sum_j \frac{\partial^2}{\partial \nu_j^2} +
O(\frac{1}{\beta^2})\right) \Delta(\frac{-\ii}{\sqrt{N \beta}} 
\frac{\partial}{\partial \nu})~,
\en
where $\Delta$ is the ordinary Vandermonde 
determinant\footnote{In deriving eq. 
(\ref{jvsdelta}) the relation $\sum_j \frac{\partial}{\partial \nu_j}=0$,
which follows from $\sum_j \nu_j=0$, has been taken into account.}. 
By operating with the r.h.s. of (\ref{jvsdelta}) we get
\eq
\label{frint6}
F_r(\beta) = \Delta(\frac{-i}{\sqrt{N \beta}} \frac{\partial}{\partial \nu})
\,\ee{-{1 \over 2} \sum_j \nu_j^2} \left[ 1  
+ \frac{N^2 - 6 N + 6}{24\beta N^2}
\sum_j \nu_j^2 + \frac{1}{24 N \beta} \sum_j \nu_j^4 + C \right]~.
\en
Finally we have to operate with $\Delta(\frac{-i}{\sqrt{N \beta}} 
\frac{\partial}{\partial \nu})$ and keep in the result only the terms
{\it quadratic in} $\nu_j^2$. In fact terms independent from $\nu_j$
are irrelevant for the reason given above, and terms of the type
$\beta^{-1} \sum_j \nu_j^4$ are of order $\beta^{-3}$ once the dependence 
of $\nu_j$ from $\beta$ is taken into account.
We shall make use of the following relations:
\eqa
\Delta(-\frac{\partial}{\partial \nu}) \ee{-\frac{1}{2} \sum_j \nu_j^2}
&=& 
\Delta(\nu) \ee{-\frac{1}{2} \sum_J \nu_j^2}~, \label{delta1} \\ 
\Delta(-\frac{\partial}{\partial \nu}) \sum_j \nu_j^2 \ee{-\frac{1}{2}
\sum_j \nu_j^2} &=&\Delta(\nu) (\sum_j \nu_j^2 - \frac{N(N-1)}{4} ) 
\ee{-\frac{1}{2} \sum_J \nu_j^2}~, \label{delta2} \\ 
\Delta(-\frac{\partial}{\partial \nu}) \sum_j \nu_j^4
\ee{-\frac{1}{2} \sum_j \nu_j^2} &=&\Delta(\nu) (\sum_j \nu_j^4 -4 
(N-\frac{3}{2}) \sum_j \nu_j^2 + C(N)) \ee{-\frac{1}{2} \sum_J \nu_j^2}~.
\nonumber\\
\label{delta3}
\ena
Eq. (\ref{delta1}) simply follows from the remark that the polynomial in 
front of the exponential is globally of order $\frac{N(N-1)}{2}$ and that it 
is completely antisymmetric under exchange of the $\nu_j$'s.
Eq. (\ref{delta2}) is a simple consequence of (\ref{delta1}): rescale
the $\nu_j^2$'s in the exponent by a parameter $\gamma$ and take on
both sides the derivative respect to $\gamma$.  
Eq. (\ref{delta3}) can be checked directly for small values
of $N$ and then proved by induction. Notice that in deriving eq. 
(\ref{delta3}) the relation $\sum_j \nu_j=0$ has been used.
By applying the above relations to eq. (\ref{frint6}) we finally obtain
\eq
\label{frint7}
F_r(\beta) = \tilde{C}(\beta,N) \Delta(\nu) 
\ee{-\frac{1}{2} \sum_j \nu_j^2 \left(1 +
\frac{1}{2 \beta N} (\frac{N}{2} - \frac{1}{N} ) + O(\beta^{-2})\right)}~.
\en
It is now sufficient to insert (\ref{frint7}) into (\ref{dierre}) and
to keep into account the definition of $\nu_j$ given in (\ref{chvar})
and the expression (\ref{casimiro}) of the quadratic 
Casimir
in order to obtain
\eq
\label{dierre2}
D_r(\beta) = \ee{-\frac{C_r}{2 N} \left[ \frac{1}{\beta} +\frac
{N^2-2}{N^2} \frac{1}{4 \beta^2} \right] + O(\beta^{-3})}~.
\en
This asymptotic behaviour can be inserted into the effective action
(\ref{zeta1}), with the rescaling of the coupling given 
in eq. (\ref{rescaling})
to determine its large $N_t$ behaviour:
\eq
\label{zetant}
\ee{S_{\rm Pol}(\beta_s=0)} = \prod_{\vec{x},i}
\sum_r  \ee{-\frac{C_r}{2 N J} \left[ 1-\frac{1}{J N_t}
(a_1 - \frac{N^2 -2}{4 N^2}) \right] +O(\frac{1}{N_t^2})}
\chi_r(P(\vec{x}+\hat\i)) \chi_r(P^\dagger(\vec{x})) ~.
\en
The leading order in $\frac{1}{N_t}$ gives, as we already knew, 
the heat-kernel action. As discussed earlier on, the 
rescaling of the coupling constant
is determined by the requirement that the r.h.s. of (\ref{zetant}) is 
$N_t$-independent up to the terms $O(\frac{1}{N_t^2})$. This fixes 
the value of $a_1$; in fact the $\frac{1}{N_t}$ term in the coefficients 
of the character expansion vanishes, for any representation, provided
\eq
\label{result}
a_1 = \frac{N^2 -2}{4 N^2}~.
\en
Let us remark that the above
result in the large $N$ limit ($a_1=1/4$) was already obtained
in \cite{bilda} by using an Eguchi--Kawai reduction scheme.

We have determined, in a purely analytic fashion, one of the two coefficients
appearing in the scaling law (\ref{rescaling}).
In the symmetric lattice ($N_t=n_t$) case eq. (\ref{rescaling}) 
describes the scaling towards the 
continuum limit ($n_t\to\infty$) of the critical temperature. 
This is the prediction that we are most
interested in, since it can be tested againts the Monte Carlo data, all
of which are performed on symmetric lattices.

For asymmetric lattices ($N_t=\rho n_t$) eq. (\ref{rescaling}) 
gives the relation between the couplings at different values of the 
asymmetry parameter\footnote{Eq. (\ref{asres}) has the same form
as eq. (\ref{rel4}); however, since we are neglecting the space-like
plaquettes, the quantum corrections $-a_1,a_1$ in (\ref{asres})
can differ from $\alpha^0_\tau,\alpha^1_\tau$ in (\ref{rel4}).} $\rho$:
\eq
\label{asres}
\beta_t(\rho n_t)=J\rho n_t + a_1 = \rho\left(\beta_t(n_t) - a_1\right) +
a_1~.
\en
The last equation relates the value of $a_1$ which we just obtained
with the parameters $\alpha_{\tau}^{0,1}$ introduced in (\ref{rel4}), 
and it gives
\eq
\alpha^1_\tau =a_1~, \hskip 1cm
\alpha^0_\tau =  - a_1~.
\label{pred}
\en

It is  interesting to compare our prediction with the results 
obtained by Karsch  ~\cite{k81} in $3+1$ dimensions and including 
space-like plaquettes. Using the same normalization, they are given by
\eq
\alpha^1_\tau =a_1=\frac{1}{4} - \frac{1}{2 N^2}~, \hskip 1cm
\alpha^0_\tau = - (0.1305 - \frac{1}{2 N^2}) = - a_1 + 0.1195~.
\en

Our prediction for the subleading term $\alpha^1_{\tau}$ coincides
with Karsch's results, while a correction, presumably due to 
contribution of the space-like plaquettes appears in $\alpha^0_{\tau}$.
\section{The leading term in the scaling law 
and comparison with the Monte Carlo data}
\label{mf}
Unlike the subleading term $a_1$ calculated in the previous section,
the  critical value $J_c$ of the rescaled coupling  $J$ depends on the
dynamics of the system and cannot be obtained
exactly, since it would require solving the model defined by
eq. (\ref{zetah1}).
An estimate of $J_c$ can be obtained
by using a mean field approximation. With the form of the effective
action given in  eq. (\ref{zetah1}) this is not an easy task, 
due to the infinite sum over the group representations appearing in eq.
(\ref{zetah1}). However the sum over the representations can be done
explicitely by means of a Poisson resummation, and the action can be recast 
in the following form (see \cite{aiz,cdmp1}):
\eqa
\label{poisson}
e^{S_{\rm Pol}^{\rm hk}(n_t=1)} &=& \prod_{\vec x,\{\hat\i=1,2\}}\Biggl\{
\left(\frac{N}{4 \pi} \right)^{1/2}
 \exp \left(\frac{N^2 - 1}{24 J} \right)\sum_P
\frac{(NJ)^{(1-N)/2}}{{\cal J}\left(\theta({\vec x})\right){\cal J}
\left(\theta(\vec x+\hat\i)\right)} \\
& & \times (-1)^{\sigma(P)  }
\sum_{\{l_{i}\}}
\exp \left[ - \frac{NJ}{2} \sum_{i = 1}^N
\left( \theta_i(\vec x) - \theta_{P(i)}(\vec x +\hat\i) + 
2 \pi l_i \right)^2 \right]\Biggl\}~,
\nonumber
\ena
where $\theta_i(\vec x)$ are the invariant angles of the Polyakov lines:
$P(\vec x)=\diag\{\ee{\ii\theta(\vec x)}\}$;  ${\cal J}(\theta)$
is the Vandermonde determinant for a unitary matrix defined in
(\ref{vandermonde}) and $P$ denotes a permutation of the indices.
The set of integers  $\{ l_i \}$ are winding numbers, related to the fact that
the angles $\theta_i(\vec x)$  are defined modulo $2 \pi$.
For special unitary groups the sum of the invariant angles must 
vanish modulo $2 \pi$; if we choose them to satisfy the relation
$\sum_i \theta_i(\vec x) = \sum_i \theta_i(\vec x + \hat\i) = 0$, 
then the relation $\sum_i l_i=0$ also follows.

An infinite sum over the integers $\{l_i\}$ also appears in 
eq. (\ref{poisson}), but
the contribution of the winding modes is known to vanish in the
deconfined phase in the large $N$ limit and it is exponentially suppressed
for finite $N$. Hence an excellent approximation is obtained 
by truncating the sum keeping only very low values of
$\{l_i\}$. 
The mean field analysis is greatly simplified in comparison with the original
form (\ref{zetah1}) of the effective action, to the extent that improvements
of the mean field approximation, obtained by considering larger and larger
clusters of ``spins'' (see Fig. 1), become viable. In this way more precise 
estimates of the critical coupling can be obtained (see \cite{id} for further
details).
\begin{figure}[b]
\begin{center}
\null\hskip -1pt
\epsfxsize = 10truecm
\epsffile{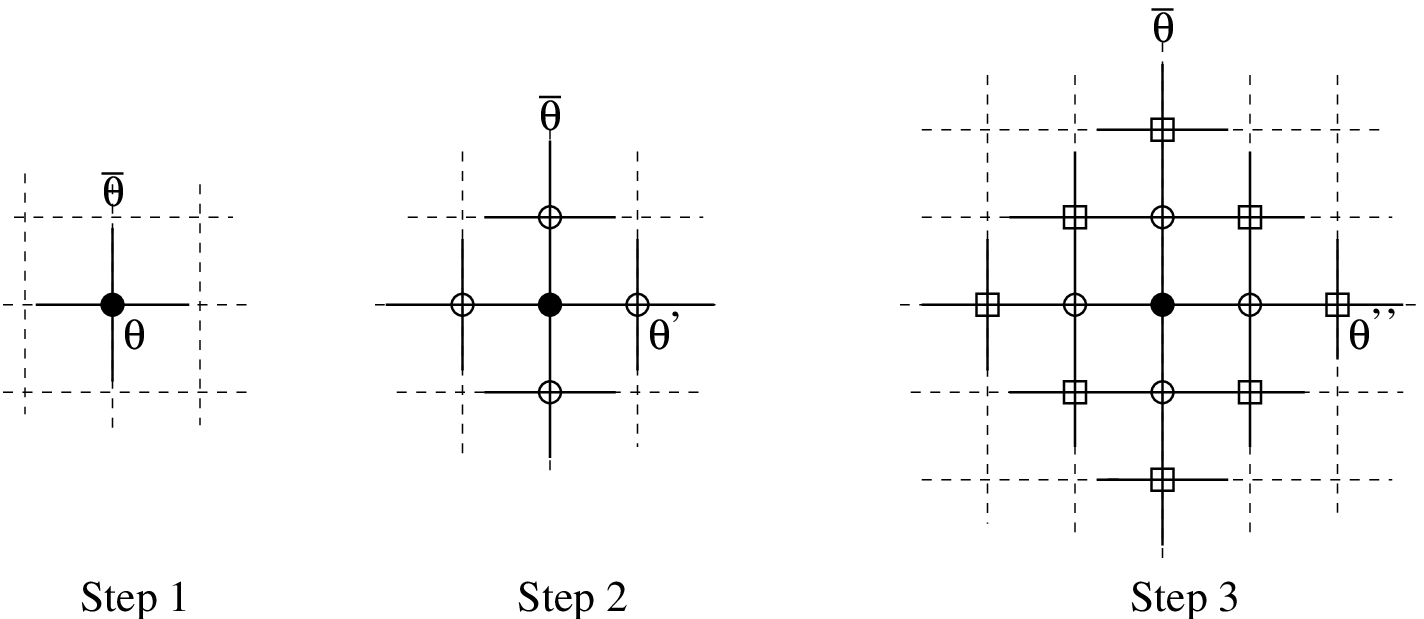}
\vskip 10pt\noindent
\end{center}
\mycaptionl{Fig. 1}{Progressively more precise Bethe-like
approximations. In {\rm Step 1} (standard mean field approximation)
we integrate only on the invariant angles $\theta_i$ of the central site,
while all the other are kept fixed to the mean value $\bar\theta$ and 
the consistency condition is $\langle P(\theta) \rangle =
P(\bar\theta)$, where $P$ is the Polyakov loop.
In the successive steps also the invariant angles $\theta_i^{\prime}$
of the nearest neibourghs ({\rm Step 2}) and $\theta_i^{\prime\prime}$
of the next to nearest neibourghs ({\rm Step 3}) are integrated over,
and the consistency condition reads $\langle P(\theta) \rangle = \langle
P(\theta^{\prime}) \rangle$. A link factor of the action (\ref{poisson})
is associated to each link drawn with a full line in the figure.}
\end{figure}
The  results of our mean field analysis  for the SU(2) and SU(3) models are 
reported in Tab. I together with the $N=\infty$ result already obtained in 
\cite{review}. Notice that the complexity of the analysis increases with 
$N$;  so  we could reach a ``step 3'' cluster in the SU(2) case, while  
we could only go as far as  a ``step 2''cluster in the SU(3) case. 
Notice however that as the number of internal degrees of freedom
increases the mean field approximation becomes more and more precise (in the 
large $N$ limit it is exact), and smaller clusters are sufficient to extract 
rather good approximations of the exact result. 
In the SU(2) case, having at our disposal
three steps we tried a (naive) extrapolation  of our estimate toward the
infinite cluster limit, assuming a simple geometric progression. This result is
reported in the fourth line of Tab. I, while in the last line we have reported,
for comparison, the Monte Carlo estimates of $J_c$ 
(that we shall discuss below) for $N=2$ and $N=3$.
\begin{table}
\mycaptionl{Table I}{Results for $J_c$  obtained by
(improved) mean field techniques  
For the $N=\infty$ case, see \cite{review}. The
boldface entries represent our best estimates of the value of $J_c$. 
In the last line we have reported the Monte Carlo estimates of $J_c$
for $N=2$ and $N=3$.}
\begin{center}
\begin{tabular}{l  c c c}
\hline\hline
\null & {$N=2$} & 
{$N=3$} & $N=\infty$ \\
\hline\hline
step 1 & 0.292 &  0.332 &  {\bf 0.351} \\
step 2 & 0.332 & {\bf 0.359} &  \\
step 3 & 0.355 & \null  &\\
extrapol. & {\bf 0.367} & \null & \\
\hline
Monte Carlo & 0.380(3) &  0.366(2) & 
\null  \\
\hline\hline
\end{tabular}
\end{center}
\end{table}
\subsection{Comparison with the Monte Carlo results}
It is interesting to compare our predictions for $J_c$ and $a_1$ with 
the existing Monte Carlo data in (2+1) dimensions. 
Very precise  estimates of the
deconfinement temperature  exist for the $N=2$ and
$N=3$ models in the range $2\leq n_t\leq 6$. These can be found 
in~\cite{t2,ctdw,eklllps} and are reported in Tab. II.
All these simulations were made with the standard Wilson
action and at $\rho=1$. 
\begin{table}
\mycaptionl{\bf Table II}{The critical coupling $\beta_c$
as  a function of the lattice size in the t direction, $n_t$, 
in the (2+1) dimensional SU(2) and SU(3) LGT, taken from~\cite{t2},
\cite{eklllps} and~\cite{ctdw}.}
\label{d2atable}
\begin{center}
\begin{tabular}{c c c }
\hline\hline 
$n_t$ & $N=2$ & $N=3$ \\  
\hline
$2$ & $\sim 0.866$ & $0.906(2)$  \\
$3$ & $\sim 1.251$ & $$           \\
$4$ & $1.630(8)$ & $1.638(6)$ \\
$6$ & $2.388(10)$ & $2.371(17)$ \\
\hline\hline
\end{tabular}
\end{center} 
\end{table}

The data show that the expected linear dependence from $n_t$ is
very well fulfilled, and, what is more important, they are precise enough
to allow to measure, besides the leading linear dependence, also the subleading
correction. 
Both in the  $N=2$ and in the $N=3$ case the Monte Carlo data fit the linear
scaling law (eq.(\ref{252}) for symmetric lattices)
\eq
\label{law}
\beta_c(n_t)=J_c n_t +  a_1
\en
with a good confidence level. If we extract from the data the best fit values of
$a_1$ and $J_c$ we obtain the results reported in Table III. The comparison with
our theoretical prediction shows a quite remarkable agreement even in the
subleading term.

Another way to compare the results of the Monte Carlo simulations with our
analytic estimates is to factor out the leading linear dependence by plotting
$\frac{\beta_c(n_t)}{n_t}$ for various values of $n_t$.
The results are shown, for both Monte Carlo and analytic data, in Table IV and
plotted in Fig. 2. The analytic data in Table IV are obtained by inserting
in (\ref{law}) our best estimate of $J_c$ and the exact result for $a_1$.
The last row in Table IV is obtained from the best fit of the Monte 
Carlo data, using (\ref{law}).
\begin{table}
\mycaptionl{Table III}{Best fit values of $J_c$ and $a_1$ 
 compared to our analytic predictions.}
\begin{center}
\begin{tabular}{  c c c c c  }
\hline\hline
\null & \multicolumn{2}{c}{$N=2$} & \multicolumn{2}{c}{$N=3$} 
 \\
\hline
\null &
Fit of M.C. data & Analytic & Fit of M.C. data & Analytic \\
\hline
$J_c$      & 0.380(3) & 0.367 & 0.366(2)  & 0.359  \\
$a_1$      & 0.106(1) & 0.125 & 0.174(6)  & 0.194  \\
\hline\hline
\end{tabular}
\end{center}
\end{table}
\begin{table}
\mycaptionl{Table IV}{MC data for $\beta_c(n_t)/n_t$ compared to our
analytic predictions obtained inserting our best estimates for 
$J_c$ and our exact result $a_1$ in eq. (\ref{252}). 
Entries in the last row are obtained by fitting the MC data (see Tab. III)}
\begin{center}
\begin{tabular}{c  c c c c c c}
\hline\hline
\null & \multicolumn{2}{c}{$N=2$} & \multicolumn{2}{c}{$N=3$} &
$N=\infty$ \\
\cline{2-6}
\raisebox{1.5ex}[0pt] {$n_t$} &
M.C. data & Analytic & M.C. data & Analytic & Analytic\\
\hline
2         & $\sim 0.433$ & 0.430 & 0.4531(8)  & 0.456 & 0.476 \\
3         & $\sim 0.417$ & 0.409 & \null      & 0.424 & 0.434 \\
4         & 0.4075(19)   & 0.398 & 0.4094(14) & 0.408 & 0.413 \\
6         & 0.3979(16)   & 0.388 & 0.3952(27) & 0.391 & 0.393 \\
\hline
$\infty$  & 0.380(3)     & 0.367 & 0.366(2)   & 0.359 & 0.351 \\
\hline\hline
\end{tabular}
\end{center}
\end{table}
\begin{figure}
\begin{center}
\null\hskip 1pt
\epsfxsize 12.5cm
\epsffile{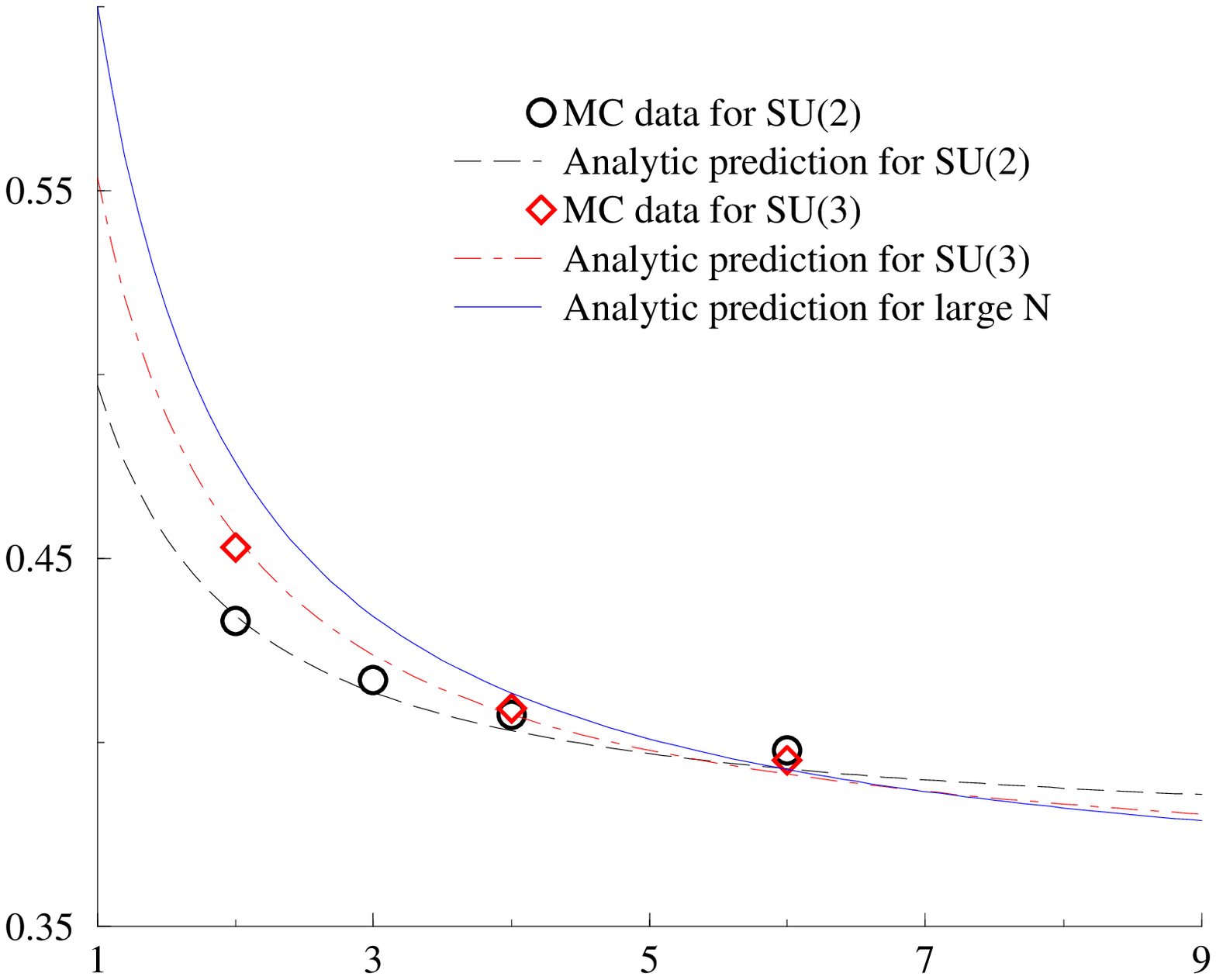}
\vskip -0.15cm\noindent
\end{center}
\mycaptionl{Fig. 2}{The avaliabe M.C data for $\beta_c(n_t)/n(t)$ are 
compared to scaling laws obtained in the present paper.} 
\end{figure}
At the end of Sec. 1 we formulated the conjecture that in $2+1$ dimensional
LGT the dynamics of the Polyakov loop is dominated by the time-like plaquettes,
and that the space-like plaquettes can therefore be neglected.
This conjecture was based solely on the fact that the time-like plaquettes
dominate for very low values of $n_t$ and that  the scaling law for the
critical coupling is the same for the theory with and without space-like
plaquettes.
The results of our analysis show that this conjecture al least predicts 
with very good approximation the location of the deconfinement transition.
The problem of whether the order of the phase transition and its critical
exponents can also be correctly predicted in the same approximation 
is an open one and it would deserve further investigation.

\vskip 0.8cm
\begin{center}
{\bf  Acknowledgements}
\end{center}
We thank S. Panzeri for many helpful discussions. 
This work was partially supported by the 
European Commission TMR programme ERBFMRX-CT96-0045.
\vfill\eject

\end{document}